# Optical pulse doublet resulting
# from the nonlinear splitting of a super-Gaussian pulse

**Christophe Finot**

*Laboratoire Interdisciplinaire Carnot de Bourgogne, UMR 6303 CNRS-Université de Bourgogne-Franche-Comté, 9 avenue Alain Savary, BP 47870, 21078 Dijon Cedex, France*

*christophe.finot@u-bourgogne.fr*

*Tel.: +33 3 80395926*

**Abstract:** We experimentally demonstrate the generation of a temporal pulse doublet from the propagation of an initial super-Gaussian waveform in a nonlinear focusing medium. The picosecond structures are characterized both in amplitude and phase and their close-to-Gaussian Fourier-transform limited shape is found in excellent agreement with numerical simulations. This nonlinear fiber-based single-stage reshaping scheme is energy efficient, can sustain GHz repetition rates and temporal compression factors around 10 are demonstrated.

**Keywords:** pulse shaping, nonlinear propagation, optical fibers.

# I. Introduction

Generation of picosecond optical pulse doublets at high repetition rates remains a technological challenge. Indeed, the optoelectronics bandwidth required to generate these double pulses is much higher than the one usually required to synthesize an isolated waveform. Therefore, all-optical approaches are fully relevant to overcome the electronics bottleneck and to provide optical doublets with a duration below 10 ps. In this context, several linear solutions have relied on photonic first-order derivation using long-period fiber gratings [1], fiber Bragg gratings [2] or on-chip microdisk resonators [3]. Nonlinear approaches in optical fibers can also be advantageous thanks to the combination of the dispersive and Kerr effects. Indeed, numerous examples of doublets have been reported in fiber lasers [4-6]. However, their repetition rates are typically a few tens of MHz. Cavity-free solutions that do not require mode-locking processes also offer attractive solutions and several strategies have been explored such as chirp-induced splitting of pulses [7], pulse copying thanks to an external triangular waveform [8, 9], parametric processes in dispersion oscillating fibers [10] or the compression of higher-order soliton pulse [11-13].

In this contribution, we introduce a new solution based on the nonlinear propagation of a super-Gaussian pulse in a fiber with anomalous dispersion. This problem, also known as the box-problem or the dam-break problem, has recently stimulated a renewed interest in optics, mainly driven by the theoretical study of the nonlinear dynamics of the temporal [14-16] or spatial [17, 18] coherent structures that emerge upon propagation. We show here that it can find practical applications. In the first part of this article, we discuss numerical simulations that demonstrate that a pair of Gaussian-like pulses emerge during the evolution of an initial flattop pulse in a single-mode waveguide with a focusing nonlinearity. We then present the experiment that validates our numerical expectations. A detailed study of the phase and intensity profiles, in the temporal and spectral domains, fully confirms our predictions and demonstrate the generation of close-to-Gaussian Fourier transform-limited doublets with a duration of 4.3 ps spaced by 15.2 ps at a repetition rate of 2 GHz.

## II. Principle and numerical simulations

We first consider the evolution of a super-Gaussian pulse propagating in an optical fiber with a Kerr nonlinear coefficient $\gamma$ and a second order anomalous dispersion $\beta_2$ ($\beta_2 < 0$). The longitudinal evolution of the temporal complex field envelope $\psi(t,z)$ in the slowly-varying envelope approximation is analytically described by the standard nonlinear Schrödinger equation (NLSE) [13]:

$$i\frac{\partial \psi}{\partial z} - \frac{\beta_2}{2}\frac{\partial^2 \psi}{\partial t^2} + \gamma |\psi|^2 \psi = 0, \qquad (1)$$

$z$ being the propagation distance and $t$ the temporal coordinate. The NLSE can be solved numerically using the split-step Fourier algorithm [13]. In order to illustrate our approach, we consider parameters typical of the standard single-mode SMF-28 fiber [19] involved in the proof-of-principle experiment described in section III. The input pulse is a third-order super-Gaussian pulse $\psi(t,0) = \sqrt{P_0} \exp\left[-\left(t/T_0\right)^6\right]$ with a full width at half maximum (fwhm) temporal duration $T_{1/2} = 2\left[\ln(2)/2\right]^{1/6} T_0 = 42$ ps and an input peak power $P_0$ of 0.9 W. They propagate in a fiber with a dispersion $\beta_2$ of -20 ps$^2$/km and a nonlinearity $\gamma$ of 1.1 /W/m.

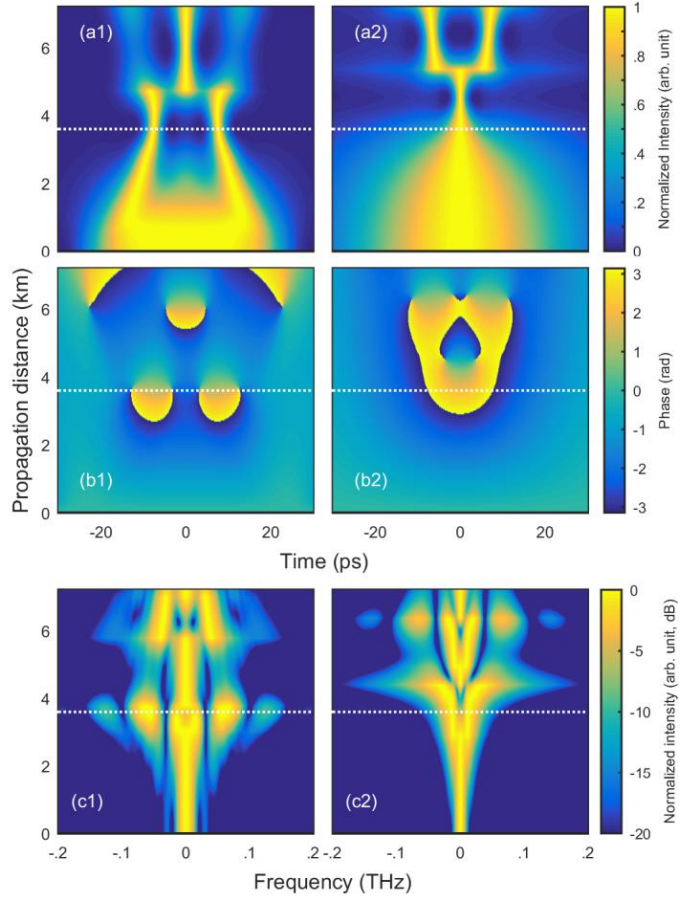

**Figure 1 :** Comparison of the longitudinal evolutions of a super-Gaussian pulse and a Gaussian waveform (panels 1 and 2 respectively). The temporal intensity and phase profiles are provided in panels a and b respectively whereas the spectral intensity profiles are given in panels c. The horizontal white dotted line represents the distance of 3.6 km. Intensities are normalized at each propagation distance by the peak intensity.

The longitudinal evolutions of the temporal intensity and phase profiles are provided in Fig. 1, panels a and b, and are compared with the dynamics of a Gaussian pulse having the same fwhm duration and peak power, $\psi'(t,0) = \sqrt{P_0} \exp(-t^2/T_0'^2)$ with $T_0' = \sqrt{2\ln(2)}\, T_{1/2}$. The evolutions of the temporal properties appear significantly different. On the one hand, the Gaussian pulse is first temporally narrowed and gets compressed to a fraction of its original width. It then splits into two substructures similarly to a higher-order soliton dynamics (the soliton number defined as $N = (\gamma P_0 T_0^2 / \beta_2)^{1/2}$ being here equal to 7.5) [12, 13]. On the other hand, the super-Gaussian pulse first experiences the emergence of two in-phase ultrashort structures in its temporal wings that then collide at the pulse center. Such a scenario should be preferred to generate a double pulse as it reduces the required fiber length and avoids the instability that affects bell-shaped pulses at their point of maximum temporal focusing [13, 20]. Details of the temporal intensity profile obtained after 3.6 km of propagation are provided in Fig. 2(a). For this distance, the super-Gaussian waveform has reshaped into a pair of well-separated and identical pulses that can be adjusted by a Gaussian profile having a fwhm duration of 2.9 ps and a temporal spacing of 15.3 ps. This stresses the significant temporal compression occuring during the nonlinear process (here a 14.5 fold compression). The energy remaining in the residual central lobe is minimal. The spectral bandwidth of the doublet (Fig. 2(b)) is significantly larger than the input pulse and is maximum after 3.6 km (see Fig. 1(c2)). The optical spectrum exhibits a modulation typical of the interference of two in-phase pulses. It can be qualitatively adjusted by a modulated Gaussian waveform with a fwhm width of 150 GHz. This leads to a time-bandwidth product of 0.44, suggesting that the Gaussian-like structures under study are rather close to the Fourier transform limit. This is in agreement with the temporal phase profile that is flat over the central part of the pulses (see Fig. 1(b1)). Note that, in the present letter, our goal is not to provide a theoretical framework of the nonlinear dynamics that is observed and we are only motivated in the qualitative trends linked to the application we demonstrate next section. Therefore, we do not explain the details of the dynamics of the box-problem and we do explicitly interpret the present results in terms of finite background solutions of the NLSE such as second-order Peregrine solitons, soliton gas or dispersive wave structures [18]. We do not carry spectral analysis in terms of nonlinear Fourier transform [21]. Let us, however, mention that the duration of the initial pulse has to be carefully chosen and that longer durations may lead to a different and more complex picture as we have experimentally described in [14] carried in a rather similar experimental configuration. Convincing explanations of the influence of the initial width can be proposed in terms of topological control [18].

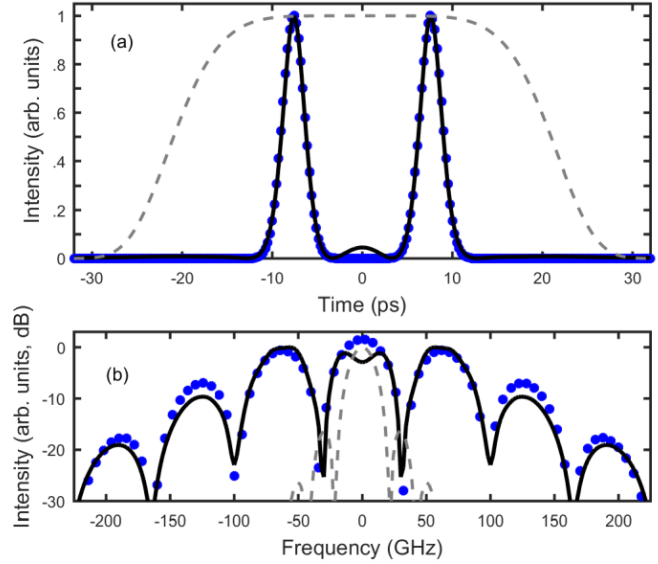

**Figure 2 :** Temporal and spectral intensity profiles (panels a and b, respectively) of the waveform after propagation in 3.6 km of fiber. Numerical simulations (solid black lines) are compared with a fit assuming a Gaussian shape of the two pulses (blue full circles). The input intensity profile is plotted with a dashed grey line.

## III. Experimental setup

In order to validate our numerical study, we have implemented an experiment based on devices that are commercially available and typical of the telecommunication industry. The experimental setup, close to the one we previously developed for the demonstration of the temporal Arago spot [22], is sketched in Fig. 3. A laser at 1550 nm emits a highly-coherent continuous wave. The initial temporal profile is obtained thanks to a Lithium Niobate intensity modulator operated at its point of null transmission and driven by an electrical pulse generator that delivers super-Gaussian pulses at a repetition rate of 2 GHz. The resulting signal is amplified using an erbium-doped fiber amplifier delivering up to 23 dBm of average power. An optical attenuator is used to control the power sent into the fiber. Particular care has been devoted so that the resulting intensity profile presents a highly symmetric profile, with a strong platitude of the high level. The corresponding optical intensity profile is plotted in Fig. 3(b) (blue solid line) and approximates the sharp edges of the ideal box waveform (red solid line). The experimental intensity profile is well fitted by a third-order super-Gaussian pulse with a fwhm duration of 42 ps. The initial optical spectrum recorded on a high-resolution optical spectrum analyzer (Fig. 3(c)) confirms the high level of symmetry of the pulse which is close to the Fourier-transform limit and the high degree of coherence of the signal.

The nonlinear propagation experiment takes place in a set of four identical 900m long optical single-mode fibers with nonlinear and dispersion parameters provided in the previous section. The high level of dispersion enables us to neglect the impact of third-order dispersion. In order to suppress the deleterious stimulated Brillouin backscattering, we have inserted three optical isolators [23]. Compared to the widely used phase modulation approach, this mitigation scheme presents the advantage of not impairing the phase of the signal under study, which is crucial for our detection stage. Indeed, the characterization of the output pulse properties is ensured by a complex optical spectrum analyzer that enables us to get access both to the spectral intensity and phase profiles. It becomes therefore possible to measure the temporal phase and intensity profiles with a temporal resolution below 1 ps.

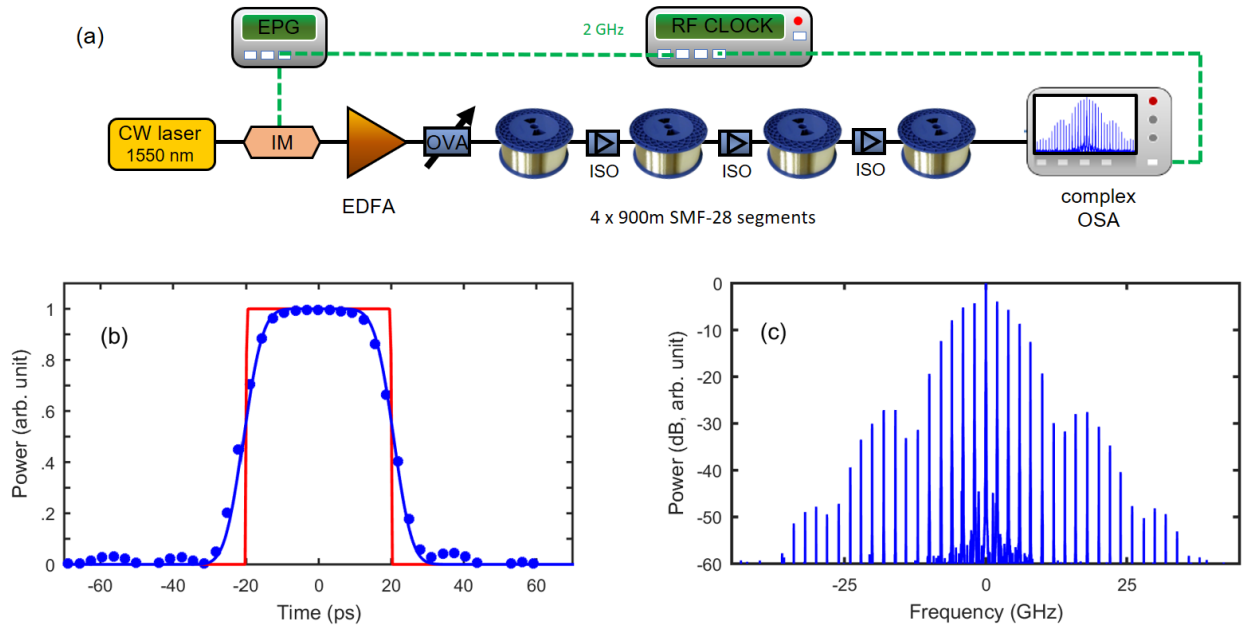

**Figure 3 :** (a) Experimental setup. CW : continuous wave, IM : intensity modulator, EPG : Electrical Pattern Generator, EDFA : Erbium Doped Fiber Amplifier, OVA : Optical Variable Attenuator, ISO : optical isolator, OSA : Optical Spectrum Analyzer. (b) Temporal intensity profile obtained after modulation. The experimental results recorded with the complex spectrum analyzer (blue circles) are compared with a fit by a third-order super-Gaussian waveform (blue solid line). The red line represents an ideal rectangular function. (c) Optical spectrum of the initial signal.

# IV. Experimental results

Instead of carrying a study in terms of propagation distance and in order to avoid cutting-back the fiber arrangement, we have measured the nonlinear dynamics as a function of the average input power. The results recorded on the optical spectrum analyzer are summarized in Fig. 4 and compared with numerical simulations of the NLSE that also take into account the distributed losses of the fiber (0.2 dB/km) as well as the lumped losses induced by the isolator (0.35 dB per isolator). The evolution of the spectrum exhibits a behavior that clearly differs from the propagation in a normally dispersive fiber where the wave-breaking evolution leads to a broadened spectrum that is not affected by any modulation and has a spectral extent that saturates [24]. The numerical predictions of the spectral profiles are found in excellent agreement with the experimental results.

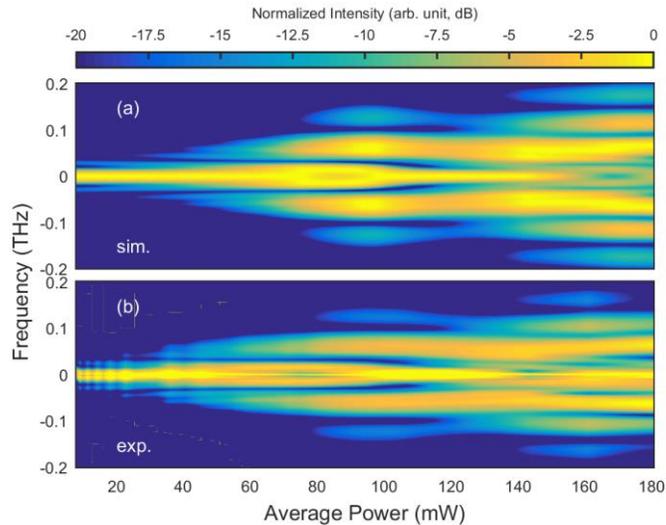

**Figure 4 :** Evolution of the optical spectrum according to the average initial power. Numerical simulations (a) are compared with experimental results (b). For clarity, the frequency comb structure of the spectrum has been removed.

The temporal intensity and phase profiles measured as a function of the average input power are provided in Fig. 5. The measurements confirm the qualitative trends observed for the longitudinal propagation: when increasing the integrated nonlinearity (by increasing the input power or the propagation length), the initial super-Gaussian waveform reshapes into two pulses. For additional propagation or nonlinearity, a third temporal component emerges from the central part of the resulting waveform. The experimental amplitude and phase features are both reproduced by numerical simulations.

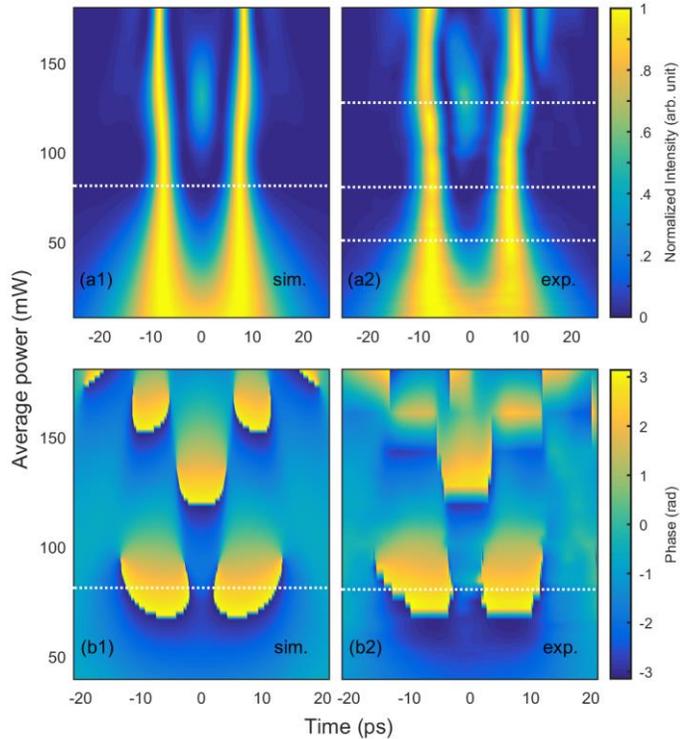

**Figure 5 :** Evolution of the output temporal intensity and phase properties (panels a and b, respectively) of the waveform as a function of the initial average power for a fixed propagation length of 3.6 km. Numerical simulations (panels 1) are compared with the experimental results (panels 2). Intensity profiles are normalized by the peak-power.

A more detailed view of the temporal profiles obtained for an average power of 80 mW (corresponding to an input peak power of 1 W and a pulse energy of 40 pJ) is proposed in Fig. 6. For this power, two identical picosecond pulses are generated and spaced by 15.2 ps. The experimental results are in remarkable agreement with numerical simulations. They confirm that the temporal phase profile over the two pulses is very flat and that the two pulsed parts are in-phase. This contrasts with other methods such as optical derivation where the leading and trailing pulses are π-phase shifted [1]. Compared to the loss-free case described in section II, the presence of losses leads to a broadening of the output pulse structures that have a fwhm duration of 4.3 ps (i.e a compression factor of 9.8 with respect to the input waveform). A beneficial impact can however be observed with the disappearance of the small bump that was predicted at the center in the ideal case.

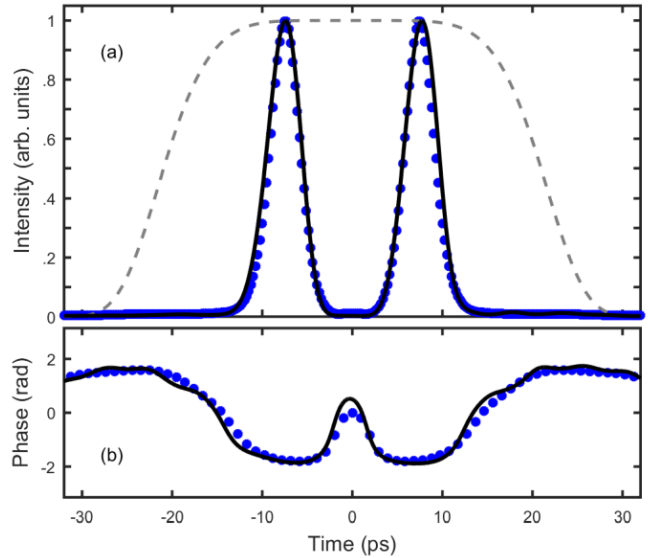

**Figure 6 :** Details of the temporal intensity and phase profiles (subplots a and b, respectively) obtained for an input average power of 80 mW. Experimental results (solid black lines) are compared with numerical simulations (blue full circles). The input intensity profile is plotted with a dashed grey line.

Finally, the temporal intensity profiles recorded for three different input powers (50, 80 and 125 mW of average power, respectively) are compared in Fig. 7. The splitting of the pulse is already apparent for pulse energies of 25 pJ. In the three cases, the intensity profiles can be adjusted with a good accuracy by the temporal superposition of two or three Gaussian structures. The delay between the two main structures is not heavily affected by the input power level.

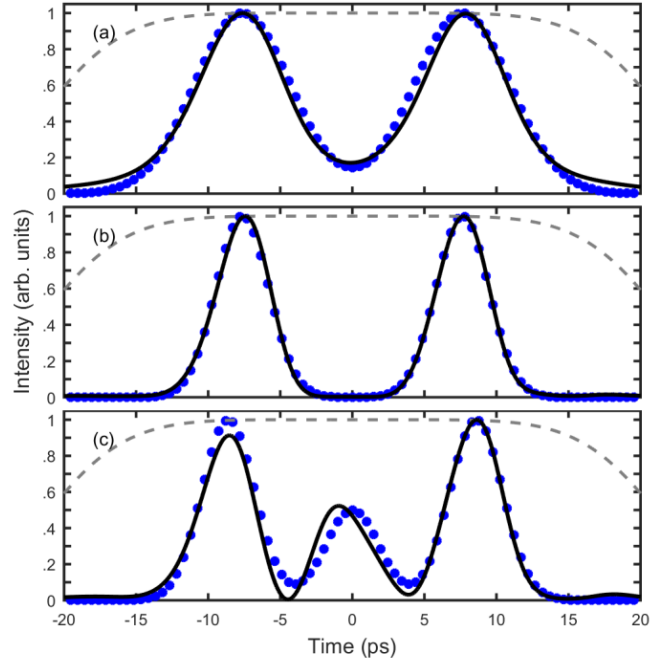

**Figure 7 :** Details of the temporal intensity profiles obtained for an input average power of 50, 80 and 125 mW (subplots a, b and c, respectively). Experimental results (solid black lines) are compared with Gaussian fits of the substructures (blue full circles). The input intensity profile is plotted with a dashed grey line.

# V. Conclusions

In conclusion, we have numerically and experimentally demonstrated the generation of a doublet of well-separated pulses resulting from the nonlinear propagation of a super-Gaussian pulse in a focusing medium. The pulses resulting from this single-stage nonlinear reshaping process are significantly shorter than the input waveform. Contrary to optical differentiation, the proposed approach is energy efficient and the losses are limited to the propagation losses. The typical powers that are required (around 1 W of input peak power) are moderate and can be further lowered using a fiber with higher nonlinearity. The approach can fully sustain GHz repetition rates.

We have focused this letter on the qualitative aspects by highlighting that the structures are in-phase and can be well described at their point of maximum focusing by a Gaussian-like profile. However, a better understanding of the nonlinear dynamics will stimulate further theoretical research with more advanced tools. Using topological perspective will enable to better determine the theoretical guidelines to apply and the balance between the duration, peak power, dispersion/diffraction and nonlinearity that has to be targeted [14, 17].

We carried this proof-of-principle demonstration using a 3.6 km long SMF-28 fiber, but nonlinear compact optical waveguides with anomalous dispersion [25] are also suitable, paving the way for chip-scale solutions of ultrashort pulse generation.

# Acknowledgements

We acknowledge the support of the Institut Universitaire de France (IUF), the Bourgogne-Franche Comté Region. CF thanks its colleagues, Frederic Audo, Bertrand Kibler, Alexandre Parriaux, Julien Fatome for fruitful discussions and technical assistance.